\documentclass[journal]{IEEEtran}
\usepackage{amsmath}
\usepackage{amssymb}
\usepackage{amsfonts}
\usepackage{graphicx}
\usepackage{epsfig}
\usepackage{subfigure}
\usepackage{psfrag}
\usepackage{cite}
\usepackage{latexsym}
\usepackage{url}
\usepackage{color}
\usepackage[compact]{titlesec}
\PassOptionsToPackage{bookmarks={false}}{hyperref}
\IEEEoverridecommandlockouts
\allowdisplaybreaks

\makeatletter
\renewcommand{\maketag@@@}[1]{\hbox{\m@th\normalsize\normalfont#1}}%
\makeatother

\begin{document}
	\title{User Grouping and Reflective Beamforming for IRS-Aided URLLC}
	
	\author{Hailiang Xie, \IEEEmembership{Graduate Student Member, IEEE,}
		Jie Xu, \IEEEmembership{Member, IEEE,}
		Ya-Feng Liu, \IEEEmembership{Senior Member, IEEE,}
		Liang Liu, \IEEEmembership{Member, IEEE,}
		and Derrick Wing Kwan Ng, \IEEEmembership{Fellow, IEEE}
		\thanks{H. Xie is with the School of Information Engineering, Guangdong University of Technology, Guangzhou 510006, China, and the Future Network of Intelligence Institute (FNii), The Chinese University of Hong Kong (Shenzhen), Shenzhen 518172, China (e-mail: hailiang.gdut@gmail.com).}
		\thanks{J. Xu is with the FNii and the School of Science and Engineering, The Chinese University of Hong Kong (Shenzhen), Shenzhen 518172, China (e-mail: xujie@cuhk.edu.cn). J. Xu is the corresponding author. }
		\thanks{Y.-F. Liu is with the State Key Laboratory of Scientific and Engineering Computing, Institute of Computational Mathematics and Scientific/Engineering Computing, Academy of Mathematics and Systems Science, Chinese Academy of Sciences, Beijing 100190, China (e-mail: yafliu@lsec.cc.ac.cn).}
		\thanks{L. Liu is with the Department of Electronic and Information Engineering, The Hong Kong Polytechnic University, Hong Kong (e-mail: liang-eie.liu@polyu.edu.hk).}
		\thanks{D. W. K. Ng is with the School of Electrical Engineering and Telecommunications, the University of New South Wales, Australia (e-mail: w.k.ng@unsw.edu.au).}}
	
	\setlength\abovedisplayskip{2.23pt}
	\setlength\belowdisplayskip{2.23pt}
	
	\maketitle
	
	\begin{abstract}
		This paper studies an intelligent reflecting surface (IRS)-aided downlink ultra-reliable and low-latency communication (URLLC) system, in which an IRS is dedicatedly deployed to assist a base station (BS) to send individual short-packet messages to multiple users. To enhance the URLLC performance, the users are divided into different groups and the messages for users in each group are encoded into a single codeword. By considering the time division multiple access (TDMA) protocol among different groups, our objective is to minimize the total latency for all users subject to their individual reliability requirements, via jointly optimizing the user grouping and block-length allocation at the BS together with the reflective beamforming at the IRS. We solve the latency minimization problem via the alternating optimization, in which the blocklengths and the reflective beamforming are optimized by using the techniques of successive convex approximation (SCA) and semi-definite relaxation (SDR), while the user grouping is updated by K-means and greedy-based methods. Numerical results show that the proposed designs can significantly reduce the communication latency, as compared to various benchmark schemes, which unveil the importance of user grouping and reflective beamforming optimization for exploiting the joint encoding gain and enhancing the worst-case user performance.
	\end{abstract}
	
	
	\newtheorem{definition}{\underline{Definition}}[section]
	\newtheorem{fact}{Fact}
	\newtheorem{assumption}{Assumption}
	\newtheorem{theorem}{\underline{Theorem}}[section]
	\newtheorem{lemma}{\underline{Lemma}}[section]
	\newtheorem{corollary}{\underline{Corollary}}[section]
	\newtheorem{proposition}{\underline{Proposition}}[section]
	\newtheorem{example}{\underline{Example}}[section]
	\newtheorem{remark}{\underline{Remark}}[section]
	\newtheorem{algorithm}{\underline{Algorithm}}[section]
	\newtheorem{proof}{Proof}[section]
	\newcommand{\mv}[1]{\mbox{\boldmath{$ #1 $}}}
	\newcounter{MYtempeqncnt}
	
	\section{Introduction}
	Ultra-reliable and low-latency communications (URLLC) has emerged as an important usage scenario for the fifth-generation (5G)-and-beyond networks to enable mission-critical Internet-of-Things (IoT) applications such as industrial automation \cite{short_packets_2016,short_packets_2018}, which require ultra-low transmission latency (e.g., less than 1 ms) and extremely high reliability (e.g., packet error probability (PEP) less than $10^{-9}$) with relatively low data rate. Different from conventional communications that rely on the Shannon capacity based on the application of sufficiently long codewords, URLLC generally relies on the short-packet transmission \cite{factory_auto}, under which the achievable rate in terms of bits per symbol can be approximated as \cite{achievable_rate} 
	\begin{small}
	\begin{align}
	\frac{d}{m} \approx \underbrace{\log_2(1+\gamma)}_{\text{Shannon capacity}}\!-\! \underbrace{\sqrt{\frac{1}{m}\left(1\!-\!\frac{1}{(1+\gamma)^2}\right)}\frac{Q^{-1}(\epsilon)}{\ln2}}_{\text{Channel dispersion}}, \label{achievable_rate}
	\end{align}
	\end{small}%
	where $\epsilon$ denotes the PEP, $d$ denotes the number of information bits, $m$ denotes the number of symbols over one code block, $\gamma$ denotes the received signal-to-noise ratio (SNR), and $Q^{-1}(x)$ denotes the inverse function of the Gaussian Q-function $Q(x)=\int_{x}^{\infty} \frac{1}{\sqrt{2\pi}}\mathrm{exp}(-t^2/2) dt$. 
	
	It is observed from (\ref{achievable_rate}) that under a given PEP $\epsilon$, the communication rate $\frac{d}{m}$ corresponds to the Shannon capacity subtracting a new channel dispersion term that is monotonically decreasing with respect to the code length $m$. In this case, to reduce the value of channel dispersion term for increasing the communication rate, it is desirable to encode different users' messages into a single codeword with an enlarged length. However, when the users' messages are jointly encoded, the Shannon capacity term in (\ref{achievable_rate}) would be determined by the minimum received SNR of these users, which may decrease as more users' messages are jointly encoded. Thus, it introduces an important but non-trivial design tradeoff in grouping users between exploiting the joint encoding gain versus avoiding the associated SNR loss. In the literature, some prior works \cite{Group1,Group2} proposed to jointly encode the messages of all users into one single codeword to minimize the channel dispersion, which, however, may lead to a severe SNR loss, especially when the channel conditions of different users become more distinctive. 

	On the other hand, intelligent reflecting surface (IRS) has been recognized as a potential key technology for beyond-5G and sixth-generation (6G) wireless networks to increase the system spectral and energy efficiency \cite{IRS1,IRS2}. In practice, IRS is a passive meta-material panel composed of a large number of reflecting elements, each of which can introduce an independent phase shift on the incident signals to reconfigure the wireless propagation environments, thus enhancing the coverage and improving the performance of worst-case users (e.g., at the cell edge)\cite{IRS1}. As a result, it is expected that IRS can play an important role for URLLC, especially for the deteriorating worst-case SNR of grouped users. In the literature, various prior works (e.g., \cite{IRS_URLLC1,IRS_URLLC2}) have been devoted to investigate the impacts of IRS on URLLC. For instance, the authors in \cite{IRS_URLLC1} considered the multi-cell multiuser orthogonal frequency division multiple access (OFDMA) URLLC systems aided by an IRS and \cite{IRS_URLLC2} studied a URLLC system with both an IRS and unmanned aerial vehicles (UAVs). However, these works did not consider the impacts of user grouping for URLLC and their results are not applicable to the system of our interest.
	
	In this letter, we study an IRS-assisted multiuser URLLC system with optimized user grouping, where a BS broadcasts individual short-packet messages to a set of distributed users assisted by an IRS. First, the users can be properly grouped such that the messages for users in each group are jointly encoded into a single codeword. Then, the time division multiple access (TDMA) protocol is employed to facilitate the downlink transmission for different user groups. Under this setup, our objective is to minimize the total latency of all users (or their total blocklength), by jointly optimizing the user grouping and the blocklength allocation at the BS, as well as the reflective beamforming at the IRS, subject to the users' individual maximum PEP and the IRS's practical reflection constraints. However, the formulated latency minimization problem is highly non-convex due to the coupling between optimization variables. To deal with this issue, we propose efficient algorithms based on alternating optimization for obtaining a high-quality suboptimal solution, in which the blocklengths and the reflective beamforming are jointly optimized by exploiting the successive convex approximation (SCA) and semi-definite relaxation (SDR) techniques, and the user grouping is updated by using the K-means and greedy-based methods. Numerical results demonstrate that by exploiting the joint encoding gain via user grouping and enhancing the worst-case user performance via IRS's reflective beamforming, the proposed designs can considerably reduce the communication latency, as compared to various conventional schemes without deploying the IRS and/or with  different users' messages encoded individually or into one single codeword. 
	
	{\it Notations:} Boldface letters refer to vectors (lower  case) or matrices (upper case). For a square matrix $\mv{S}$, ${\mathtt{Tr}}(\mv{S})$ denotes its trace, and $\mv{S}\succeq \mv{0}$ means that $\mv{S}$ is positive semidefinite. For an arbitrary-size matrix $\mv{M}$, ${\mathrm{rank}}(\mv{M})$ and $\mv{M}^H$ denote its rank and conjugate transpose, respectively. The distribution of a circularly symmetric complex Gaussian (CSCG) random vector with mean vector $\mv{x}$ and covariance matrix $\mv{\Sigma}$ is denoted by $\mathcal{CN}(\mv{x,\Sigma})$; and $\sim$ stands for ``distributed as''. $\mathbb{C}^{x\times y}$ denotes the space of $x\times y$ complex matrices. $\mathbb{Z}^{+}$ denotes the set of positive integers. $\|\cdot\|_2$ denotes the Euclidean norm of a vector.
	
	\section{System Model and Problem Formulation}
	
	We consider an IRS-aided multi-group downlink URLLC system, where one BS sends short-packet messages to $K$ users, assisted by one IRS. It is assumed that the BS and users are single-antenna devices\footnote{Notice that the single-antenna BS is assumed in this work for revealing the fundamental limits of user grouping and reflective beamforming for multiuser URLLC. Nevertheless, the design principles herein can also be extended to more general scenarios with multiple antennas at the BS, in which the transmit beamforming can be employed jointly for further performance enhancement.}, and the IRS is equipped with $N$ reflecting elements. Let $\mathcal{K}\triangleq\{1,\ldots,K\}$ denote the set of users. Let $h_k\in\mathbb{C}$, $\mv{g}\in\mathbb{C}^{N\times1}$, and $\mv{f}_k\in\mathbb{C}^{N\times1}$ denote the channels from the BS to user $k$, from the BS to the IRS, and the IRS to user $k\in\mathcal{K}$, respectively. We consider a quasi-static flat-fading channel model, where the wireless channels remain unchanged within each transmission block of our interest, but may vary over different blocks. Furthermore, we assume that the perfect channel state information (CSI) is available at the BS (via channel estimation methods in, e.g., \cite{IRS_est_OFDM}) for resource allocation design.
	
	Suppose that the $K$ users are assigned into $G$ groups, denoted by set $\mathcal{G} \triangleq \{1,\ldots,G\}$. Let $\mathcal{K}_i$ denote the set of users in group $i\in\mathcal{G}$, and $|\mathcal{K}_i|$ denote its cardinality. Also, each user is assigned into only one group for reducing the transmission latency, and accordingly we have $\mathcal{K}_i\cap\mathcal{K}_j=\emptyset, \forall i, j\in\mathcal{G}, i\neq j$, and $\cup_{i\in\mathcal{G}}\mathcal{K}_i=\mathcal{K}$. Let $d_k$ denote the number of information bits that need to be conveyed to user $k$. As a result, there are in total $D_i=\sum_{k\in\mathcal{K}_i}d_k$ bits for group $i$. The BS then jointly encodes the $D_i$-bit information for group $i$ into a single packet (codeword) with length of $m_i$ symbols, where $m_i\in \mathbb{Z}^+$.
	
	Next, the BS broadcasts the encoded packets to the $G$ groups in a TDMA manner to avoid severe inter-group interference, where the communication block is separated into $G$ slots, each for one group. In each slot $i\in \mathcal{G}$, let $s_i$ denote the transmitted signal from the BS to user group $i$, where $s_i,\forall i$, are assumed to be independent and identically distributed (i.i.d.) CSCG random variables with zero mean and unit variance, i.e., $s_i\sim\mathcal{CN}(0,1)$. On the other hand, at the IRS, let $\mv v=[e^{j\theta_{1}},\ldots,e^{j\theta_{N}}]^H$ denote the reflective beamforming vector, where $\theta_n\!\in\![0, 2\pi)$ denotes the phase shift imposed by the $n$-th reflecting element. Notice that the reflective beamforming vector at the IRS is assumed to remain unchanged over the $G$ time slots. This is due to the fact that adaptively reconfiguring the phase shifters at the IRS for every time slot may consume extra time, signaling overhead, and energy, and thus may not be able to be implemented at the interested time scale for URLLC. Accordingly, we have the cascaded end-to-end channel from the BS to user $k$ as $h_k+\mv v^H\mv\phi_k$, where $\mv\phi_k = \mathrm{diag}(\mv{f}^H_{k})\mv{g}\in\mathbb{C}^{N\times1}$. In this case, the signal received by user $k$ in group $i$ is accordingly expressed as
	\begin{small}
	\begin{align}
	y_k = \sqrt{P}(h_k+\mv v^H\mv\phi_k)s_i+ z_k, k\in\mathcal{K}_i, i\in\mathcal G, \label{receive_signal}
	\end{align}
	\end{small}%
	where $P$ is the constant maximum transmit power at the BS and $z_k$ denotes the additive white Gaussian noise (AWGN) at user $k$ with zero mean and variance $\sigma_k^2$, i.e., $z_{k}\sim\mathcal{CN}(0,\sigma_k^2), \forall k\in\mathcal K$. Accordingly, the received SNR at user $k$ is given by
	\begin{small}
	\begin{align}
	\gamma_k = P|h_k+\mv v^H\mv\phi_k|^2/\sigma_k^2.\label{SNR}
	\end{align}
	\end{small}%
	Since the achievable rate of each user group is limited by the worst-case user with the minimum received SNR, we denote the minimum SNR of all users in group $i$ as
	\begin{small}
	\begin{align}
	\gamma^{\min}_i=\min_{k\in\mathcal{K}_i} \{\gamma_k\}.\label{min_SNR}
	\end{align}
	\end{small}%
	According to the achievable rate formula with finite blocklength in (\ref{achievable_rate}), the worst-case PEP of users in group $i$ can be written as a function of blocklength $m_i$ and SNR $\gamma^{\min}_i$, i.e.,
	\begin{small}
	\begin{align}
	\epsilon(m_i, \gamma^{\min}_i)\!=\! Q\left(\frac{m_i\ln(1\!+\!\gamma^{\min}_i)\!-\!\ln2D_i}{\sqrt{m_i}\sqrt{1\!-\!(1\!+\!\gamma^{\min}_i)^{-2}}}\right), i\in\mathcal{G}. \label{PEP}
	\end{align}
	\end{small}%
	Similarly, when $\epsilon_i\le0.5$, the blocklength of group $i$ can be written as a function of $\epsilon_i$ and $\gamma^{\min}_i$, i.e.,
	\begin{small}
	\begin{align}
	m(\epsilon_i,\gamma^{\min}_i)\!=\! \frac{D_i\ln2}{\ln(1\!+\!\gamma^{\min}_i)}\!+\!\frac{\lambda_i^2}{2}\!+\!\lambda_i\sqrt{\!\left(\frac{\!\lambda_i}{2}\!\right)^{\!2}\!\!\!+\!\frac{D_i\ln2}{\ln(1\!+\!\gamma^{\min}_i)}}, \label{BL_function}
	\end{align}
	\end{small}%
	where
	\begin{small}
	\begin{align}
	\lambda_i = \sqrt{1-(1+\gamma^{\min}_i)^{-2}}\,Q^{-1}(\epsilon_i)/\ln(1+\gamma^{\min}_i). \label{lambda}
	\end{align}
	\end{small}%

	Our objective is to minimize the total latency of the system (or equivalently the users' total blocklength) by jointly optimizing the user grouping\footnote{Note that the number of user groups, $G$, also needs to be optimized, as implied in $\{\mathcal{K}_i\}$.} and the reflective beamforming at the IRS as well as the blocklength of each group, subject to the practical constraints on the users' maximum PEP at each group and the IRS reflection. The latency minimization problem is formulated as 
	\begin{small}
	\begin{align}
	&\mathrm{(P1):}\mathop\mathtt{min}_{\{\mathcal{K}_i\},\{m_i\},\mv{v}}~\sum_{i\in\mathcal{G}}m_i \nonumber \\
	&{\mathtt{s.t.}}\hspace{0.2cm} Q\left(\!\frac{m_i\ln(1\!+\!\gamma_k)\!-\!\ln2D_i}{\sqrt{m_i\left[1\!-\!(1\!+\!\gamma_k)^{-2}\right]}}\!\right)\!\le\!\epsilon_{\max}, \forall k\!\in\!\mathcal{K}_i, i\!\in\!\mathcal{G}, \label{Problem:ori:1} \\
	&\hspace{0.7cm}|v_n|= 1, \forall n\in\{1,\ldots,N\}, \label{Problem:ori:2} \\
	&\hspace{0.7cm}\mathcal{K}_i\cap\mathcal{K}_j=\emptyset,\cup_{i\in\mathcal{G}}\mathcal{K}_i=\mathcal{K}, \forall i, j\in\mathcal{G}, i\ne j, \label{Problem:ori:3} \\
	&\hspace{0.7cm}m_i\in\mathbb{Z}^+, \forall i\in\mathcal{G}. \label{Problem:ori:4}
	\end{align}
	\end{small}%
	Problem (P1) is highly non-convex and challenging to be optimally solved. To tackle this difficulty, we propose efficient algorithms based on the alternating optimization to obtain an suboptimal solution to problem (P1). In particular, we first optimize $\{m_i,\mv{v}\}$ under given $\{\mathcal{K}_i\}$ in Section III and then optimize the user grouping $\{\mathcal{K}_i\}$ in Section IV.
	
	\section{Joint Blocklength and Reflective Beamforming Design}
	In this section, we propose an efficient algorithm to jointly optimize the blocklength $\{m_i\}$ and the reflective beamforming $\mv{v}$ under any given user grouping $\{\mathcal{K}_i\}$. 
	
	\subsection{Problem Reformulation}
	With given $\{\mathcal{K}_i\}$, we relax the positive integer constraints in (\ref{Problem:ori:4})\footnote{After obtaining the fractional solution of $\{m_i\}$ in (P2), we can round them up to the nearest integer to obtain a feasible solution for the original problem (P1), where the enlarged blocklength results in a smaller PEP such that the constraints in (\ref{Problem:Joint1:1}) are satisfied.} as the continuous one and recast problem (P1) as
	\begin{small}
	\begin{align}
	&\mathrm{(P2):}\mathop\mathtt{min}_{\{m_i\},\mv{v}}
	~\sum_{i\in\mathcal{G}}m_i \nonumber \\
	&{\mathtt{s.t.}} Q\left(\frac{m_i\ln(1\!+\!\gamma_k)\!-\!\ln2D_i}{\sqrt{m_i\left[1\!-\!(1\!+\!\gamma_k)^{-2}\right]}}\right)\!\le\!\epsilon_{\max}, \forall k\!\in\!\mathcal{K}_i, i\!\in\!\mathcal{G}, \label{Problem:Joint1:1} \\
	&\hspace{0.6cm}|v_n|= 1, \forall n\in\{1,\ldots,N\} \label{Problem:Joint1:2}.
	\end{align}
	\end{small}%
	However, problem (P2) is still non-convex. To facilitate the derivation, we first reformulate constraint (\ref{Problem:Joint1:1}) as
	\begin{small}
	\begin{align}
	&\mathcal{R}\left(m_i, \gamma_k\right)=m_i\!\ln(1\!+\!\gamma_k)\!-\!\ln2D_i\nonumber \\
	&~~-\!Q^{-1}(\epsilon_{\max})\sqrt{m_i\!\left[1\!-\!(1\!+\!\gamma_k)^{-2}\right]}\!\ge\!0,\forall k\!\in\!\mathcal{K}_i, i\!\in\!\mathcal{G},\label{cons_PEP}
	\end{align}
	\end{small}%
	and then introduce auxiliary optimization variables $\{\mu_k\}$, where $\mu_k$ denotes the lower bound of SNR at user $k$. Accordingly, problem (P2) can be rewritten as the following equivalent problem:
	\begin{small}
	\begin{align}
	&\mathrm{(P2.1):}\mathop\mathtt{min}_{\{m_i\},\mv{v},\{\mu_k\}}~\sum_{i\in\mathcal{G}}m_i \nonumber \\
	&\hspace{1.5cm}{\mathtt{s.t.}}\hspace{0.3cm}
	\mathcal{R}\left(m_i, \mu_k\right)\ge 0, \forall k\!\in\!\mathcal{K}_i, i\!\in\!\mathcal{G}, \label{Problem:Joint2:1} \\
	&\hspace{2.4cm}P|h_k\!+\!\mv v^H\!\mv\phi_k|^2\!\ge\!\mu_k\sigma_k^2, \forall k\!\in\!\mathcal{K}_i, i\!\in\!\mathcal{G}, \label{Problem:Joint2:2} \\
	&\hspace{2.4cm}|v_n|= 1, \forall n\in\{1,\ldots,N\}. \label{Problem:Joint2:3}
	\end{align}
	\end{small}%
	Next, we apply the SDR technique to convexify the non-convex constraints in (\ref{Problem:Joint2:2}) and (\ref{Problem:Joint2:3}). To this end, we first define $|h_k+\mv v^H\mv\phi_k|^2=\bar{\mv{v}}^H\mv{R}_k\bar{\mv{v}}+|h_k|^2$, where
	\begin{small}
	\begin{align}
	{\mv R}_k = \begin{bmatrix}
	\mv\phi_k\mv\phi_k^H & \mv\phi_k h_k^H \\
	h_k\mv\phi_k^H & 0
	\end{bmatrix} ~\mathrm{and} ~\bar{\mv v} = \begin{bmatrix}
	\mv{v} \\ 1 \end{bmatrix}. \label{R_v}
	\end{align}
	\end{small}%
	Then, we define $\mv{V}=\bar{\mv{v}}\bar{\mv{v}}^H$ with $\mv{V}\succeq\mv{0}$ and rank$(\mv{V})\le1$. Motivated by the idea of SDR, we relax the non-convex rank-one constraint on $\mv{V}$ and obtain a relaxed version of problem (P2.1) as
	\begin{small}
	\begin{align}
	&\mathrm{(P2.2):}\mathop\mathtt{min}_{\{m_i\},\mv{V},\{\mu_k\}}\sum_{i\in\mathcal{G}}m_i \nonumber \\
	&\hspace{0.3cm}{\mathtt{s.t.}}\hspace{0.3cm}\mathcal{R}\left(m_i, \mu_k\right)\ge 0, \forall k\!\in\!\mathcal{K}_i, i\!\in\!\mathcal{G}, \label{Problem:Joint3:1} \\
	&\hspace{1.2cm}\mathrm{Tr}(\mv{R}_k\mv{V})+|h_k|^2\ge \mu_k\sigma_k^2/P, \forall k\in\mathcal{K}_i, i\in\mathcal{G}, \label{Problem:Joint3:2} \\
	&\hspace{1.2cm}\mv{V}\!\succeq\!\mv{0}, \mv{V}_{n,n}\!=\!1, \forall n\!\in\!\{1,...,N\!+\!1\}. \label{Problem:Joint3:3}
	\end{align}
	\end{small}%
	However, problem (P2.2) is still challenging to be optimally solved due to the non-convex constraints in (\ref{Problem:Joint3:1}). In the next subsection, we solve (P2.2) by updating the optimization variables $\{m_i, \mu_k, \mv{V}\}$ iteratively via the SCA technique. Notice that the obtained solution to problem (P2.2) may not be feasible for problem (P2.1) (i.e., the SDR may not be tight). Therefore, after solving (P2.2), a Gaussian randomization procedure \cite{IRS_multiuser, IRS_Maxmin} should be further adopted to recover a rank-one solution to (P2.1). In general, the Gaussian randomization process needs to be implemented multiple times and the best solution among them is selected
	as the solution to (P2.1).
	
	\subsection{Proposed Solution to (P2.2)}
	Now, we focus on solving problem (P2.2) via SCA in an iterative manner. First, consider a particular iteration $l\ge 1$. At the current point $\{m_i^{(l-1)}, \mu_k^{(l-1)}, \mv{V}^{(l-1)}\}$, we establish a lower bound of non-convex function $\mathcal{R}\left(m_i, \mu_k\right)$ in (\ref{cons_PEP}) by replacing the non-convex component by its first-order Taylor expansion with respect to $m_i$ and $\mu_k$, as shown in (\ref{Taylor}) at the top of next page.
	
	\begin{figure*}[!t]
		\vspace{-0.3cm}
		\setcounter{MYtempeqncnt}{\value{equation}}
		\setcounter{equation}{\value{equation}}
		\begin{small} 
		\begin{flalign}
		\mathcal{R}\left(m_i, \mu_k\right) \ge
		~& \mathcal{R}\left(m_i^{(l-1)}, \mu_k^{(l-1)}\right) +\left[\ln(1+\mu_k^{(l-1)})-Q^{-1}(\epsilon_{\max})\sqrt{1-(1+\mu_k^{(l-1)})^{-2}}/2\sqrt{m_i^{(l-1)}}\right]\left(m_i-m_i^{(l-1)}\right)\nonumber \\
		& +\left\{m_i^{(l-1)}/\ln(1+\mu_k^{(l-1)})-Q^{-1}(\epsilon_{\max})\sqrt{m_i^{(l-1)}}/\left[\sqrt{(1+\mu_k^{(l-1)})^{2}-1}\left(1+\mu_k^{(l-1)}\right)^2\right]\right\}\left(\mu_k-\mu_k^{(l-1)}\right)\nonumber \\
		\triangleq~& \hat{\mathcal{R}}\left(m_i, \mu_k~|~m_i^{(l-1)}, \mu_k^{(l-1)}\right), \forall k\in\mathcal{K}_i, i\in\mathcal{G}. \label{Taylor}
		\end{flalign}
		\end{small}%
		\hrulefill
		\vspace{-0.5cm}
	\end{figure*}

	Furthermore, due to the non-convex constraint in (\ref{Problem:Joint3:1}), a trust region constraint is introduced to tighten the lower bound in (\ref{Taylor}) \cite{SCP,TR}. Let $\mv{m}=[m_1,\dots,m_G]$ and $\mv{\mu}=[\mu_1,\dots,\mu_K]$ denote the vectors consisting of the optimization variables $\{m_i\}$ and $\{\mu_k\}$, respectively. Accordingly, the imposed trust region constraint is expressed as
	\begin{align}
	\Vert[\mv{m},\mv{\mu}]^T-[\mv{m}^{(l-1)},\mv{\mu}^{(l-1)}]^T \Vert_2\le\Gamma, \label{Trush}
	\end{align} 
	where $\Gamma$ is the trust region radius. By adding the trust region constraint (\ref{Trush}) and replacing $\mathcal{R}\left(m_i, \mu_k\right)$ by $\hat{\mathcal{R}}(m_i, \mu_k|m_i^{(l-1)}, \mu_k^{(l-1)})$ in (\ref{Taylor}), problem (P2.2) is approximated as
	\begin{small}
	\begin{flalign}
	&\mathrm{(P2.3):}\mathop\mathtt{min}_{\{m_i\},\mv{V},\{\mu_k\}}\sum_{i\in\mathcal{G}}m_i \nonumber \\
	&\hspace{0.7cm}{\mathtt{s.t.}}\hspace{0.1cm}\hat{\mathcal{R}}(m_i, \mu_k|m_i^{(l-1)}, \mu_k^{(l-1)})\!\ge\!0, \forall k\!\in\!\mathcal{K}_i, i\!\in\!\mathcal{G}, \label{Problem:Joint4:1} \\
	&\hspace{1.4cm}\Vert[\mv{m},\mv{\mu}]^T-[\mv{m}^{(l-1)},\mv{\mu}^{(l-1)}]^T\Vert_2\le\Gamma, \label{Problem:Joint4:2}\\
	&\hspace{1.4cm}(\ref{Problem:Joint3:2})\mathrm{~and~}(\ref{Problem:Joint3:3}).\nonumber
	\end{flalign}
	\end{small}%
	Problem (P2.3) is convex and thus can be solved optimally by CVX \cite{CVX}. Let $\{m_i^{\star}, \mu_k^{\star}, \mv{V}^{\star}\}$ denote the obtained optimal solution of problem (P2.3).
	
	Next, we update $\{m_i\}$. Notice that the PEP is strictly decreasing with respect to the blocklength $m_i$ and the received SNR $\gamma^{\min}_i$ \cite{ULDL}, respectively. As such, the optimal blocklength $m_i$ for group $i$ must ensure that $\epsilon(m_i, \gamma^{\min}_i)=\epsilon_{\max}$, which, however, may not hold at the solution to (P2.3) due to the approximation. To deal with this issue, in each iteration $l$ of SCA, after solving problem (P2.3), we substitute the obtained $m(\epsilon_{\max},\min_{k\in\mathcal{K}_i}\mu_k^{\star})$ into the objective function of (P2.3). If the objective value decreases, then we update the blocklength $m_i$ as $m_i^{(l)}=m(\epsilon_{\max},\min_{k\in\mathcal{K}_i}\mu_k^{\star})$ and accordingly set $\mv{V}^{(l)}=\mv{V}^{\star}$; otherwise, we reduce the trust region $\Gamma$ and solve problem (P2.3) again until $\Gamma$ is less than a given threshold. By implementing this, we obtain a high-quality solution to problem (P2.2)

	\section{User Grouping}
	In this section, we design the user grouping $\{\mathcal{K}_i\}$ with any given reflective beamforming $\mv{v}$, to fully exploit the benefit of joint encoding. As exhaustively searching the optimal user grouping solution requires a prohibitively large computational complexity, we propose two low-complexity user grouping schemes based on the K-means and greedy-based clustering, respectively. Notice that by alternately implementing the solution to (P2.1) and the proposed user grouping algorithms, we can obtain an efficient solution to problem (P1).
	
	%Furthermore, as the solution of the proposed user grouping algorithms may not be optimal, the proposed alternating optimization with K-means clustering and with greedy clustering do not guarantee to decrease monotonically. We thus terminate the alternating optimization until the objective value no longer decreases.

	\subsection{User Grouping with K-means Clustering}
	In this subsection, we propose a user grouping scheme by using the K-means clustering \cite{K_means}, in which the users with proximate SNR values are classified into the same group. This scheme is motivated by the fact that the communication performance users in each group is limited by that with the minimum SNR, and it is thus desired to separate the high-SNR users from low-SNR ones for minimizing the total latency. 
	
	In particular, under given $\mv{v}$, the SNR of each user $k$ is expressed as $\gamma_k$ in (\ref{SNR}). Suppose that the $K$ users are partitioned into $G$ groups, each of which is associated with a (virtual) cluster SNR center, which is defined as the mean of the users' SNR values. In the K-means clustering scheme, we first choose $G$ users with most distinct SNR values and each of which is assigned into one of the $G$ user groups. Next, we perform the following two-step iteration. In the first step, we calculate the distance between each user's SNR value and each cluster center, and accordingly assign each user to the group with the nearest cluster center. In the second step, we update the value of each cluster center as the mean of all users' SNR values in that group. The two steps are iteratively implemented until the cluster centers do not change.
	
	Notice that for K-means clustering, how to properly choose the value of $G$ is critical. In this scheme, we implement the above procedures for any $G\in\{1, \ldots, K\}$, and choose $G$ achieving the best performance as the optimal number of groups. Also notice that the complexity of the K-means clustering based user grouping scheme is $\mathcal{O}(K^3)$.

%	\begin{table}[htp]
%		\begin{center}
%			\hrule
%			\vspace{0.1cm}\textbf{Algorithm 1}: User Grouping with K-means Clustering \vspace{0.1cm}
%			\hrule \vspace{0.1cm}
%			\begin{itemize}
%				\item[1:] \textbf{Initialize:} $G=0, L_k=0, k\in\mathcal{K}, \mathcal{G}=\{1,\ldots,G\}$ and $\gamma_k=\gamma_k(\mv{v})$;
%				\item[2:] \textbf{For} $l\in\mathcal{K}$ \textbf{do}
%				\item[3:] \hspace{0.3cm} $G=G+1$;
%				\item[4:] \hspace{0.3cm} Choose $G$ users' SNR as an initial $G$ centers $\mathcal{C}=\{c_1,\dots,c_G\}$;
%				\item[5:] \hspace{0.3cm} Set these $G$ users belong to $G$ groups, i.e., $\{\mathcal{K}^l_1,\dots,\mathcal{K}^l_G\}$;
%				\item[6:] \hspace{0.3cm} For $k\in\mathcal{K}$, set $\mathcal{K}^l_i =\mathcal{K}^l_i\cup\{k\}$, where $i=\arg\min_{i\in\mathcal{G}}|\gamma_k-c_i|$;
%				\item[7:] \hspace{0.3cm} For $i\in\mathcal{G}$, recalculate $c_i=\sum_{k\in\mathcal{K}^l_i}\gamma_k/|\mathcal{K}^l_i|$;
%				%			\item[8:] \hspace{0.3cm} Set $\mathcal{K}^l_i=\emptyset, \forall i\in\mathcal{G}$ if $\mathcal{C}$ unchanges;
%				\item[8:] \hspace{0.3cm} Repeat Steps 6 and 7 until $\mathcal{C}$ no longer changes;
%				\item[9:] \hspace{0.3cm} Calculate the total latency $L_l=\sum_{i\in\mathcal{G}}m_i$ by (\ref{BL_function}) with $\{\mathcal{K}^l_i\}$;
%				\item[10:] \textbf{End for}
%				\item[11:] \textbf{Return:} the best solution $\{\mathcal{K}^\star_i\}=\{\mathcal{K}^l_i\}$, where $l=\arg\min_{l\in\mathcal{K}}L_l$.
%			\end{itemize}
%			\vspace{0.1cm} 
%			\hrule \label{Table:1}
%			\vspace{-0.73cm} 
%		\end{center}
%	\end{table}

	\subsection{User Grouping with Greedy-based Clustering}
	This subsection proposes another heuristic user grouping scheme, based on the greedy method \cite{Greedy} that is implemented in an iterative manner as follows. We denote $\bar{\mathcal{K}}$ as the set of un-grouped users, $\mathcal{K}_1, \mathcal{K}_2, \ldots, \mathcal{K}_G$ as the set of users in the $G$ groups, and $\mathcal{K}^{(l)}$ as the set of grouped users in each iteration $l\ge 1$, where $\mathcal{K}^{(0)}$ is an empty set. To start with, we initialize the $\bar{\mathcal{K}} = \mathcal{K}$ and $G = 0$. In each iteration $l\ge1$, we temporarily assign any one user $k\in \bar{\mathcal{K}}$ into one existing group $\mathcal{K}_i$ for any $i\in \{1, \ldots, G\}$, or a new group $\mathcal{K}_{G+1} = \{k\}$, and calculate the corresponding achieved total latency by the grouped users in $\mathcal{K}^{(l-1)}$ and the newly added user $k$ (in group $i$ or new group $i=G+1$) as $m^{(l)}(k,i), \forall k\in\bar{\mathcal{K}}, i\in\{1,\dots,G,G+1\}$ via (\ref{BL_function}). Let $(k^{(l)}, i^{(l)}) = \arg\max_{k\in\bar{\mathcal{K}}, i\in\{1,\dots,G,G+1\}} m^{(l)}(k,i)$. Then if $i^{(l)}\le G$, we assign user $k^{(l)}$ in group $\mathcal{K}_{i^{(l)}}$ in this iteration, i.e., $\mathcal{K}_{i^{(l)}} = \mathcal{K}_{i^{(l)}} \cup \{k^{(l)}\}$; otherwise, assign user $k^{(l)}$ into a new group, i.e., $\mathcal{K}_{G+1} = \{k^{(l)}\}$ and set $G \gets G+1$. Accordingly, we update $\bar{\mathcal{K}} =  \bar{\mathcal{K}}\setminus \{k^{(l)}\}$ and $\mathcal{K}^{(l)} =  \mathcal K^{(l-1)} \cup \{k^{(l)}\}$. The above iteration will be implemented $K$ times until all the users are grouped. Note that the computational complexity of the user grouping with greedy-based clustering is $\mathcal{O}(K^3)$ for the worst case, which is the same as that for K-means clustering.
	
%	\begin{table}[htp]
%		\begin{center}
%			\hrule
%			\vspace{0.1cm}\textbf{Algorithm 2}: User Grouping with Greedy-based Clustering \vspace{0.1cm}
%			\hrule \vspace{0.1cm}
%			\begin{itemize}
%				\item[1:] \textbf{Initialize:} $G=0, \bar{\mathcal{K}}=\mathcal{K}$, and 
%				$\mathcal{K}_i=\emptyset,\forall i=1,\dots,G+1$;
%				\item[2:] \textbf{While} $\bar{\mathcal{K}}\ne\emptyset$ \textbf{do}
%				\item[3:] \hspace{0.3cm} \textbf{For} $k\in\bar{\mathcal{K}}$ \textbf{do}
%				\item[4:] \hspace{0.7cm} \textbf{For} $l=1,\dots,G+1$ \textbf{do}
%				\item[5:] \hspace{1.1cm} $\mathcal{K}^l_i = \mathcal{K}_i\cup\{k\}$, $\mathcal{K}^l_j = \mathcal{K}_j, \forall i\ne j, i, j=1,\dots,G+1$;
%				\item[6:] \hspace{1.1cm} Calculate the total latency $\mathcal{F}(\{\mathcal{K}^l_i\},\epsilon_{\max})$ via (\ref{BL_function});
%				\item[7:] \hspace{0.7cm} \textbf{End for}
%				\item[8:] \hspace{0.7cm} $\{\mathcal{K}^k_i\}=\arg\min_{\{\mathcal{K}^l_i\}}\mathcal{F}(\{\mathcal{K}^l_i\},\epsilon_{\max})$;
%				\item[9:] \hspace{0.3cm} \textbf{End for}
%				\item[10:] \hspace{0.3cm} $k^{\star}=\arg\min_{k}\mathcal{F}(\{\mathcal{K}^k_i\},\epsilon_{\max})$;
%				\item[11:] \hspace{0.3cm} Set $\bar{\mathcal{K}} = \bar{\mathcal{K}}/{k^\star}$ and $ \{\mathcal{K}_i\}=\{\mathcal{K}^{k^\star}_i\}$;
%				\item[12:] \hspace{0.3cm} $G=G+1$ if $\mathcal{K}^{k^\star}_{G+1}\ne\emptyset$.
%				\item[13:] \textbf{End while}
%				\item[14:] \textbf{Return:} the user grouping $\{\mathcal{K}_i\}$.
%			\end{itemize}
%			\vspace{0.1cm} \hrule \label{Table:2}
%		\end{center}
%	\end{table}

	\begin{figure*}[htb]
	\begin{minipage}[t]{0.32\linewidth}
		\includegraphics[width=5.63cm]{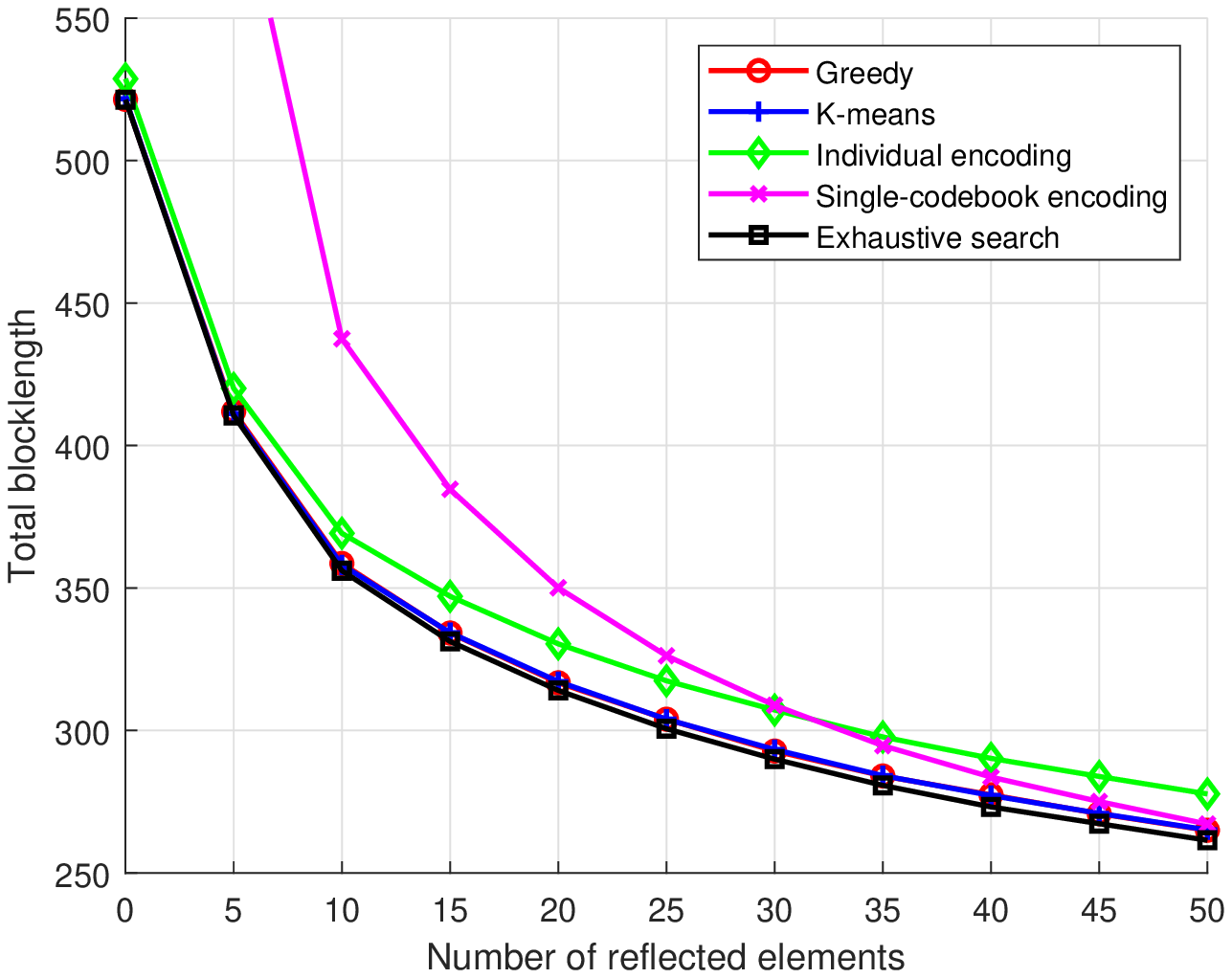}
		\caption{The total blocklength versus the number of reflecting elements $N$ at the IRS when $K=5$.}\label{fig:NumUnits}
	\end{minipage}
	\begin{minipage}[t]{0.32\linewidth}
		\centering
		\includegraphics[width=5.63cm]{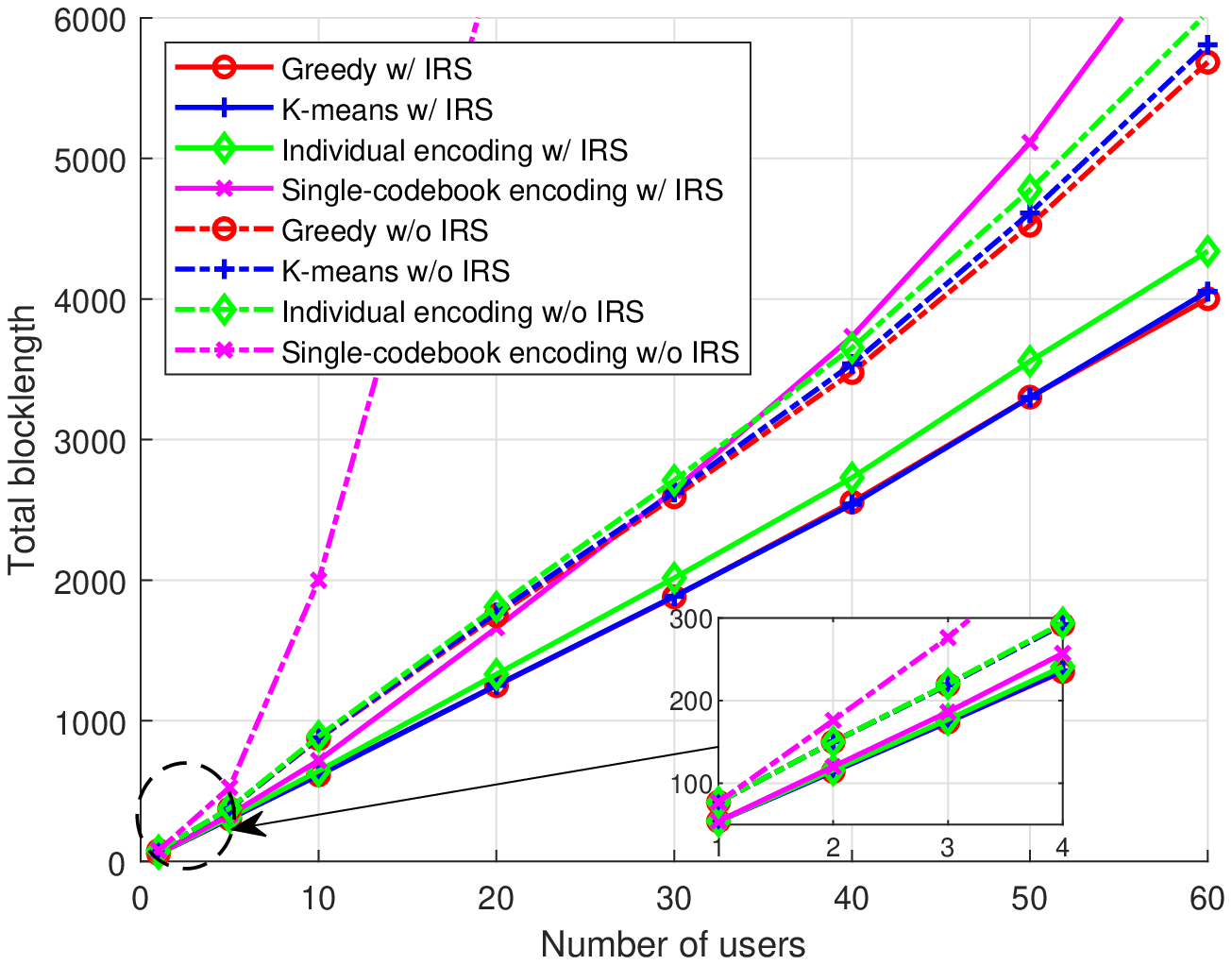}
		\caption{The total blocklength versus the number of users $K$ when $N=20$.}
		\label{fig:NumUsers}
	\end{minipage}
	\begin{minipage}[t]{0.32\linewidth}
		\centering
		\includegraphics[width=5.63cm]{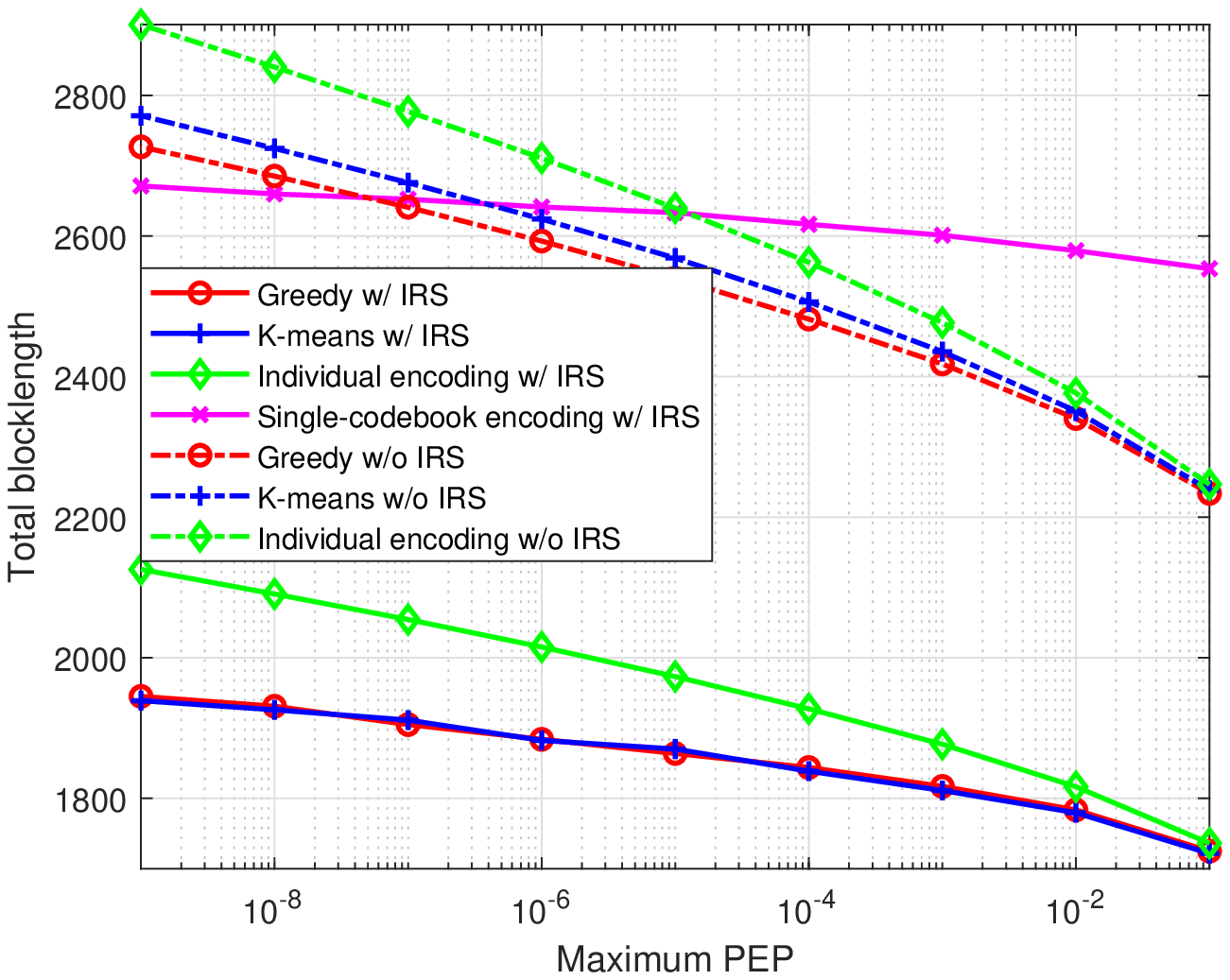}
		\caption{The total blocklength versus the maximum PEP $\epsilon_{\max}$ when $N=20$ and $K=30$.}
		\label{fig:MaxPEP}
	\end{minipage}
	\centering
	\vspace{-0.5cm}
	\end{figure*}
	
	\section{Numerical Results}
	This section provides numerical results to validate the performance of our proposed designs for IRS-aided URLLC. In the simulation, we consider a Cartesian coordinate system, where a BS is located at $(0,0)$, an IRS is deployed at $(100~\text{m}, 20~\text{m})$, and $K$ users are randomly located in a circle with radius of $10~\text{m}$ and centre of $(100~\text{m},0)$. We assume that the information bits conveyed to each user are identical to be $32$ bytes \cite{3GPP}. The noise power at each user $k$ is set as $\sigma^2_k=-80~{\text{dBm}}, \forall k\in\mathcal K$. We consider Rayleigh fading for both the BS-user and IRS-user links and LOS channel for the BS-IRS link, with path-loss exponents being $3.5$, $2.5$, and $2.0$, respectively \cite{IRS_multiuser, IRS_Maxmin}. All the results are averaged over multiple independent channel realizations.
	
	For comparison, we consider various benchmark schemes, e.g., without IRS (i.e., $N=0$), and/or with the individual encoding (i.e., $G = K$ such that each user's message is encoded individually) and single-codebook encoding (i.e., $G=1$ such that all the $K$ users' messages are jointly encoded into one single codeword), respectively. Besides, we also consider the exhaustive search based user grouping to achieve the performance upper bound or latency lower bound, in which we first solve problem (P2) with any given $\{\mathcal{K}_i\}$, and then compare the correspondingly obtained latency values to find the minimum one.
	
	Fig. \ref{fig:NumUnits} shows the total blocklength (latency) versus the number of reflecting elements $N$ at the IRS when $K=5$. It is observed that the performance of our proposed designs (with K-means and greedy) approaches closely to the upper bound achieved by the exhaustive search and significantly outperform other benchmark schemes. It is also observed that the individual encoding scheme outperforms the single-codebook encoding when $N<30$, and the opposite is true when $N>30$, while the single-codebook encoding design performs close to our proposed designs for large $N$. This shows that equipping more IRS elements $N$ is beneficial in improving the SNR of the worst-case users. Furthermore, it is observed that there is a marginal gain in the latency reduction with respect to the increasing $N$, as large $N$ will no longer be a limiting factor to the performance in the considered setup when the number of users is small.
	
	Fig. \ref{fig:NumUsers} shows the total blocklength versus the number of users $K$, where $N=20$ is set for the schemes with IRS. It is observed that for both cases with and without IRS, our proposed designs with user grouping considerably outperform the benchmark schemes with individual encoding and single-codebook encoding. In fact, the performance gap increases with the number of users as the increasing number of users impose more stringent constraints to problem (P1). This justifies the benefits of grouping users for joint encoding. It is also observed that the greedy-based user grouping design outperforms the K-means-based one in the case without IRS. This is due to the fact that although the K-means-based user grouping design minimizes the intra-group SNR variance, it may produce more groups than that of the greedy-based one.
	
	Fig. \ref{fig:MaxPEP} shows the total blocklength versus the maximum PEP when $N=20$ and $K=30$. Similar observations are made as in Fig. \ref{fig:NumUsers}. Furthermore, it is observed that the individual encoding scheme performs close to the proposed designs when the PEP becomes large, while the proposed user grouping designs significantly outperform the benchmark schemes when the PEP becomes small.
	
	\section{Conclusion}
	In this letter, we considered an IRS-aided multiuser URLLC system with user grouping, where a BS broadcasts short-packet messages to grouped users with the help of the IRS. We minimized the total latency, by jointly optimizing the BS's user grouping, the blocklength of different groups, and the IRS's reflective beamforming. By using the optimization and clustering techniques, efficient algorithms were proposed to obtain an efficient suboptimal solution to the formulated latency minimization problem. Numerical results showed the superior performance of the proposed designs over existing baseline schemes. This demonstrates the benefits of joint encoding and IRS in enhancing the multiuser URLLC performance.


\begin{thebibliography}{99}

		\bibitem{short_packets_2016} G. Durisi, T. Koch, and P. Popovski, ``Toward massive, ultrareliable, and low-latency wireless communication with short packets,'' {\it Proc. IEEE}, vol. 104, no. 9, pp. 1711--1726, Aug. 2016.
		
		\bibitem{short_packets_2018} P. Popovski, J. J. Nielsen, C. Stefanovic, E. de Carvalho, E. Ström, K. F. Trillingsgaard, A-S. Bana, D. M. Kim, R. Kotaba, J. Park, and R. B. Sørensen, ``Wireless access for ultra-reliable low-latency communication: Principles and building blocks,'' {\it IEEE Network}, vol. 32, no. 2, pp. 16--23, Apr. 2018.
		
%		\bibitem{short_packets_2017} P. Schulz, M. Matthe, H. Klessig, M. Simsek, G. Fettweis, J. Ansari, S. A. Ashraf, B. Almeroth, J. Voigt, I. Riedel, A. Puschmann, A. Mitschele-Thiel, M. Muller, T. Elste, and M. Windisch, ``Latency critical IoT applications in 5G: Perspective on the design of radio interface and network architecture,'' {\it IEEE Commun. Mag}, vol. 55, no. 2, pp. 70--78, Feb. 2017.
		
		\bibitem{factory_auto} O. N. C. Yilmaz, Y.-P. E. Wang, N. A. Johansson, N. Brahmi, S. A. Ashraf, and J. Sachs, ``Analysis of ultra-reliable and low-latency 5G communication for a factory automation use case,'' in {\it Proc. IEEE ICCW}, Jun. 2015, pp. 1190--1195.
		
		\bibitem{achievable_rate} Y. Polyanskiy, H. V. Poor, and S. Verdu, ``Channel coding rate in the finite blocklength regime,'' {\it IEEE Trans. Inf. Theory}, vol. 56, no. 5, pp. 2307--2359, May 2010.
		
		\bibitem{Group1} K. F. Trillingsgaard and P. Popovski, ``Downlink transmission of short packets: Framing and control information revisited,'' {\it IEEE Trans. Commun.}, vol. 65, no. 5, pp. 2048--2061, May 2017.
		
		\bibitem{Group2} D. Tuninetti, B. Smida, N. Devroye, and H. Seferoglu, ``Scheduling on the Gaussian broadcast channel with hard deadlines,'' in {\it Proc. IEEE ICC},  May 2018, pp. 1--7.
		
		\bibitem{IRS1} Q. Wu and R. Zhang, ``Towards smart and reconfigurable environment: Intelligent reflecting surface aided wireless network,''
		{\it IEEE Commun. Mag.}, vol. 58, no. 1, pp. 106--112, Jan. 2020.
		
		\bibitem{IRS2} Ö. Özdogan, E. Björnson, and E. G. Larsson, ``Intelligent reflecting surfaces: Physics, propagation, and pathloss modeling,"  {\it IEEE Wireless Commun. Lett.}, vol. 9, no. 5, pp. 581--585, May 2020.

		\bibitem{IRS_URLLC1} W. R. Ghanem, V. Jamali, and R. Schober, ``Joint beamforming and phase shift optimization for multicell IRS-aided OFDMA-URLLC systems,'' 2020. [Online]. Available: https://arxiv.org/abs/2010.07698.
		
		\bibitem{IRS_URLLC2} A. Ranjha and G. Kaddoum, ``URLLC facilitated by mobile UAV relay and RIS: A joint design of passive beamforming, blocklength and UAV positioning,'' {\it IEEE Internet Things J.}, vol. 8, no. 6, pp. 4618--4627, Mar. 2021.
		
		\bibitem{IRS_est_OFDM} Y. Yang, B. Zheng, S. Zhang, and R. Zhang, ``Intelligent reflecting surface meets OFDM: Protocol design and rate maximization,'' {\it IEEE Trans. Commun.}, vol. 68, no. 7, pp. 4522--4535, Jul. 2020.
		
		\bibitem{IRS_multiuser} Q. Wu and R. Zhang, ``Intelligent reflecting surface enhanced wireless network via joint active and passive beamforming," {\it IEEE Trans. Wireless Commun.}, vol. 18, no. 11, pp. 5394--5409, Nov. 2019.
		 
		\bibitem{IRS_Maxmin} H. Xie, J. Xu, and Y.-F. Liu, ``Max-min fairness in IRS-aided multi-cell MISO systems with joint transmit and reflective beamforming," {\it IEEE Trans. Wireless Commun.}, vol. 20, no. 2, pp. 1379--1393, Feb. 2021.
		
		\bibitem{SCP} S. Boyd, ``Sequential convex programming, lecture notes for EE364b: Convex Optimization II, Stanford University," 2011. [Online]. Available: http://www.stanford.edu/class/ee364b/lectures.html
		
		\bibitem{TR} A. R. Conn, N. I. M. Gould, and P. L. Toint, {\it Trust-Region Methods}. Philadelphia, PA, USA: SIAM, 2000.

		\bibitem{CVX} M. Grant and S. Boyd, ``CVX: MATLAB software for disciplined convex programming,'' 2016. [Online] Available: {\url{http://cvxr.com/cvx}}
		
		\bibitem{ULDL} C. Shen, T. Chang, H. Xu, and Y. Zhao, ``Joint uplink and downlink transmission design for URLLC using finite blocklength codes,'' in {\it Proc. IEEE ISWCS}, Aug. 2018, pp. 1--5.
		
		%\bibitem{Blocklength}Y. Hu, M. Serror, K. Wehrle, and J. Gross, ``Finite blocklength performance of cooperative multi-terminal wireless industrial networks,'' in {\it IEEE Trans. Veh Technol.}, vol. 67, no. 7, pp. 5778--5792, Jul. 2018.

		\bibitem{K_means} D. Arthur and S. Vassilvitskii,  ``K-means++: The advantages of careful seeding'' in {\it Proc. Symp. Discrete Algorithms,} 2007, pp. 1027--1035. 
		
		\bibitem{Greedy} T. H. Cormen, C. E. Leiserson, R. L. Rivest, and C. Stein, {\it Introduction to Algorithms}. Cambridge, MA, USA: MIT Press, 2009.
		
		\bibitem{3GPP} 3GPP, ``3GPP tr 38.913 v15.0.0: Study on scenarios and requirements for next generation access technologies; (release 15),'' Tech. Rep., Jun. 2018.
		
		
		
	\end{thebibliography}
\end{document}